\documentstyle[a4,a4wide,12pt,epsfig]{article}
\def\Jo#1#2#3#4{{#1} {\bf #2}, #3 (#4)}

\def\NPB{{\em Nucl. Phys.} {\bf B}}

\def\PLB{{\em Phys. Lett.}  {\bf B}}
\def\PRL{\em Phys. Rev. Lett.}
\def\PRD{{\em Phys. Rev.} {\bf D}}

\def\PR{{\em Phys. Rev. }}
\def\PRP{{\em Phys. Rep. }}

\def\ra{\rightarrow}
\def\be{\begin{equation}}
\def\ee{\end{equation}}

\def\gs{\mathrel{
   \rlap{\raise 0.511ex \hbox{$>$}}{\lower 0.511ex \hbox{$\sim$}}}}
\def\ls{\mathrel{
   \rlap{\raise 0.511ex \hbox{$<$}}{\lower 0.511ex \hbox{$\sim$}}}}

\newcommand{\onbb}{neutrinoless double beta decay }

\newcommand{\bnmu}{\mbox{$\bar{\nu}_\mu$} }

\newcommand{\nel}{\mbox{$\nu_e$} }
\newcommand{\nmu}{\mbox{$\nu_\mu$} }

\newcommand{\emm}{\mbox{$\langle m_{\mu \mu} \rangle$ }}
\newcommand{\cs}{cross section}

\newcommand{\Slash}[1]{\mbox{$#1\hspace{-.6em}/$}}
\newcommand{\ba}{\begin{array}{c}}
\newcommand{\baz}{\begin{array}{cc}}
\newcommand{\bad}{\begin{array}{ccc}}
\newcommand{\bea}{\begin{equation} \begin{array}{c}}
\newcommand{\eea}{ \end{array} \end{equation}}
\newcommand{\ea}{\end{array}}
\newcommand{\D}{\displaystyle}

\newcommand{\nnu}{\mbox{$0\nu \beta \beta$ }}

\newcommand{\mab}{\mbox{$\langle m_{\alpha \beta} \rangle $}}
\newcommand{\mmm}{\mbox{$\langle m_{\mu \mu} \rangle$}}

\newcommand{\mee}{\mbox{$\langle m_{ee} \rangle$}}

\newcommand{\meu}{\mbox{$\langle m_{e \mu} \rangle$}}
\newcommand{\omem}{\mbox{$\overline{m_{e \mu}}$}}
\begin{document}
\title{\hfill { \bf {\normalsize DO--TH 99/13}}\\ 
\hfill { \bf {\normalsize hep-ph/9907203}}\\ \vskip 1.5cm
\bf Trimuon production in $\nu$N--scattering as a probe of 
massive neutrinos}
\author{M. Flanz$^a$
\footnote{Email address: flanz@dilbert.physik.uni-dortmund.de} 
$\;$, W. Rodejohann$^a$
\footnote{Email address: rodejoha@dilbert.physik.uni-dortmund.de}$\;$, 
K. Zuber$^b$
\footnote{Email address: zuber@physik.uni-dortmund.de}\\
{\it \normalsize $^a$Lehrstuhl f\"ur Theoretische Physik III,}\\ 
{\it \normalsize $^b$Lehrstuhl f\"ur Experimentelle Physik IV,}\\
{\it \normalsize Universit\"at Dortmund, Otto--Hahn Str. 4,}\\ 
{\it \normalsize 44221 Dortmund, Germany}}
\date{}
\maketitle
\thispagestyle{empty}
\begin{abstract}
The lepton--number violating process $\nmu N \ra \mu^- \mu^+ \mu^+ X$ is 
studied for the first time in connection
with Majorana neutrino masses of the second generation. 
The sensitivity for light and heavy Majorana neutrinos is investigated. 
The ratio with respect to the standard model charged current 
process is improved by some orders of 
magnitude if compared to previously discussed 
Majorana induced $\Delta L_{\mu} = 2$ 
processes. 
Non--observation of this process in previous experiments 
allows to demand the effective mass 
to be $\langle m_{\mu \mu} \rangle \ls 10^{4}$ GeV, 
being more stringent than previously discussed direct bounds, however 
still unnaturally high. 
Therefore, in the forseeable future, 
indirect bounds on effective masses other than \mee{} will 
be more stringent. 
\end{abstract}

\newpage
\section{Introduction}
Investigation of lepton--number violating processes is one of the most
promising ways of probing physics beyond the standard model. 
A particular aspect of this topic is lepton--number violation 
in the neutrino sector, which in the case of massive neutrinos
would allow a variety of new phenomena \cite{zuber}. This emerges
immediately in case
of Majorana masses of the neutrinos, which are predicted in 
most GUT--theories \cite{GUT}. 
For \nel the searches for Majorana neutrinos mainly rely on
\onbb (0$\nu\beta\beta$), resulting in an upper limit on the 
effective Majorana mass  
$\langle m_{ee} \rangle = \mid \sum U_{em}^2 m_m \eta^{\rm CP}_m \mid $ 
of about 0.2 eV \cite{hdmo},
where $m_m$ are the mass eigenvalues, $\eta^{\rm CP}_m = \pm 1$ 
the relative CP--phases and $U_{em}$ the mixing matrix
elements. In general, there is a $3 \times 3$ matrix of effective 
Majorana masses, the elements being 
\bea \label{meffmatrix}
\mab = |(U \, {\rm diag}(m_1 \eta^{\rm CP}_1
, m_2 \eta^{\rm CP}_2 , m_3\eta^{\rm CP}_3) 
U^{\rm T})_{\alpha \beta}| \\[0.3cm]
= \left| \D \sum m_m \eta^{\rm CP}_m U_{\alpha m} U_{\beta m} \right|
\mbox{ with }  \alpha, \beta = e , \, \mu  , \, \tau .     
\eea
In this paper we explore the possibility to learn 
about Majorana masses associated with the second
generation. 
The process under study is muon lepton--number violating 
($\Delta L_{\mu} = 2$) 
trimuon production in neutrino--nucleon scattering via charged
current reactions (CC)
\be
\label{proces}
\nmu N \ra \mu^- \mu^+ \mu^+ X . 
\ee
The muonic analogy for the quantity measured in \onbb reads 
$\langle m_{\mu \mu} \rangle  =
 \mid \sum U_{\mu m}^2 m_m \eta^{\rm CP}_m \mid $ and 
is investigated for both light and heavy Majorana neutrinos. 
The relevant diagram is shown in Fig.\ \ref{feyndia}, which 
also defines the kinematics. 
Alternative ways discussed in the literature to obtain direct 
information about \emm are
muon capture on nuclei \cite{moha} and 
lepton number violating K--decays like $K^-\ra \pi^+ \mu^- \mu^-$
\cite{halprin,ng,abad,littenberg}.
The experimental knowledge of effective Majorana masses other 
than the one measured in \nnu allows only rather poor limits.
The best values obtained are from 
muon--positron conversion in sulfur (therefore sensitive to 
$\langle m_{\mu e} \rangle^2$)  
and lepton--number violating 
$K$--decays: 
\be
\bad
\frac{\D \sigma (^{32}{\rm S} + \mu^- \ra  ^{32}{\rm Si}^{\ast} + e^+)}
{\D \sigma ( ^{32}{\rm S} + \mu^- \ra  ^{32}{\rm P}^{\ast} + \nu_{\mu})} 
< 9 \cdot 10^{-10} & \Rightarrow & \langle m_{\mu e} \rangle < 
\left\{ \ba 0.4 \mbox{ GeV (singlet)} \\[0.2cm] 
1.9 \mbox{ GeV (triplet)} \ea \right.  \\[0.5cm] 
\frac{\D \Gamma (K^- \ra \pi^+ \mu^- \mu^-)}{\D \Gamma (K^- \ra {\rm all})} 
< 1.5 \cdot 10^{-4} & \Rightarrow & \langle m_{\mu \mu} \rangle < 
1.3 \cdot  10^5 \; {\rm GeV} . 
\ea
\ee
Here the experimental limits are taken from the PDG 
\cite{PDG} and for the mass bounds 
the theoretical results given 
in \cite{doi} and \cite{abad} are used (all ratios are 
proportional to $\langle m_{\mu \alpha} \rangle^2 $). 
For muon--positron conversion two results are given, depending on whether 
the proton pairs in the final state are in a spin singlet or triplet state, 
respectively. To our knowledge, there are no direct limits on other 
elements of $\langle m_{\alpha \beta} \rangle$. 
Note that we are considering {\it direct} limits, i.\ e.\ 
measuring processes which are directly dependent on the respective 
quantity without making any further assumptions. We will show that 
using the scattering process (\ref{proces}) instead of the rare $K$ 
decay allows to set a limit better by one order of 
magnitude on $\langle m_{\mu \mu} \rangle$, 
which is however still too high to give a physical 
mass matrix in Eq.\ (\ref{meffmatrix}).   
{\it Indirect} bounds, e.\ g.\  using unitarity of the 
mixing matrix and oscillation experiments 
will of course be much more stringent.   
Direct production of Majorana neutrinos heavier than 100 GeV has 
been studied for various collider types   
($e^- e^+ , \; e \mu, \; pp, \; p \overline{p} , \; e^- p$) 
\cite{collider,cvetic} with typical results 
of a few to some hundred events per year for high--energy and luminosity 
machines.

\section{Model and calculation}
\begin{figure}[hp]
\setlength{\unitlength}{1cm}
\begin{center}
\epsfig{file=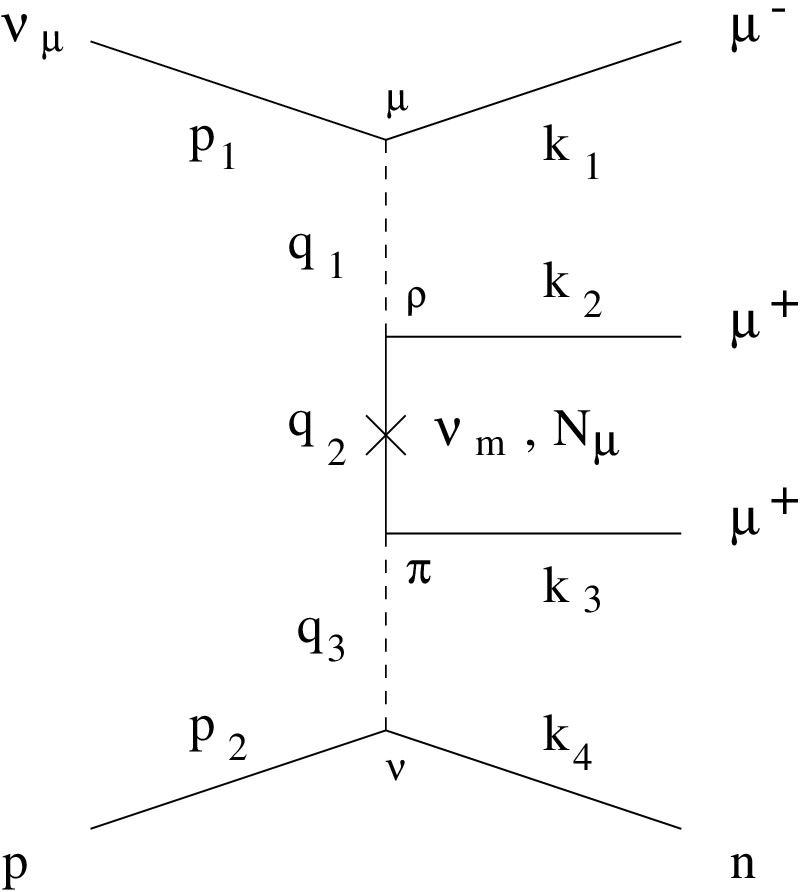,width=13cm,height=10cm}
\end{center}
\caption{\label{feyndia}Feynman diagram for the considered process.
It is $q_2 = q_1 - k_2 = p_1 - k_1 - k_2$. For the crossed diagram
$k_2 \mbox{ and }k_3$ are exchanged and we denote the corresponding momentum
of the Majorana neutrino with $\tilde{q}_2 = q_1 - k_3 = p_1 - k_1 - k_3$.
For the $W$ momenta holds: $q_1 = p_1 - k_1$ and $q_3 = k_4 - p_2$.}
\end{figure}

Using the diagram shown in Fig.\ \ref{feyndia}  
plus its crossed version we get for the squared invariant amplitude 
three terms, each factorizing nicely in three parts. 
At the upper and lower vertex we have the standard V--A term. 
The contribution of the Majorana neutrino, i.\ e.\ the 
part $ W^+ W^+ \to \mu^+ \mu^+$ 
is well known from the 
theory of $\nnu \! \!$. 
For the calculation one might follow the strategy in 
Kayser's textbook \cite{kayser} or use the Feynman rules 
from \cite{feyrul}, we will do the former.   
From here on we refer to this part of the diagram 
as the ``$\nnu \! \!$--like'' process. 
We use for particles the standard Lagrangian 
\be
\mbox{$\cal L$} = - \frac{\D g}{\D \sqrt{2}}
\D \sum_m U_{\mu m} \overline{\mu} \gamma_{\alpha} \gamma_-
U_{\mu m} \nu_m W^{\alpha}  
\ee  
where $\gamma_{\pm} = \frac{1}{2} (1 \pm \gamma_5)$. 
We denote the Majorana neutrino with $\nu_m$, the muon with $\mu$ 
and $U_{\mu m}$ is an element of the unitary matrix connecting weak 
interaction eigenstates with mass eigenstates.   
For the $\nnu \! \!$--like contribution 
to the first matrix element we have: 
\be
\mbox{$\cal{M}$}_1 \propto \left[ 
\overline{\nu_m} \gamma_{\rho} \gamma_+ U_{\mu m}^{\ast} \mu \right] 
\left[
\overline{\nu_m} \gamma_{\pi} \gamma_+ U_{\mu m}^{\ast} \mu \right] . 
\ee
To bring this in a form suitable for inserting the fermion propagator 
we use the relation: 
\be 
\overline{\nu_m} \gamma_{\rho} \gamma_+ \mu = 
- \overline{\mu^c}  \gamma_{\rho} \gamma_- \nu_m^c . 
\ee
Here $\mu^c$ means the charge conjugated spinor of the muon. 
For a given spinor $\psi$ charge conjugation has the properties:   
\be
\baz 
\psi^c = C \overline{\psi}^T , \;& 
\overline{\psi^c} = - \psi^T C^{-1} \\[0.2cm]
C^{-1} \gamma_{\mu} C = - \gamma_{\mu}^T , \; & 
C^{-1} \gamma_5 C = \gamma_5^T = \gamma_5 . 
\ea
\ee
In the standard Dirac notation $C = i \gamma_2 \gamma_0$ is the 
charge conjugation matrix. Since $\nu_m$ is a Majorana particle it 
has the property 
$\nu_m^c = \lambda_m^{\ast} \nu_m$, $\lambda_m$ 
being a phase factor in the field expansion of $\nu_m$ connected with 
the intrinsic CP parity $\eta^{\rm CP}_m$, see e.\ g.\ \cite{kayser}. 
For the expansion in terms of spinors and creation/annihilation 
operators the following relations are valid: 
\be
\baz 
\nu_m \propto f u + \lambda_m f^+ v , \; &
\mu   \propto f u + \overline{f}^+ v , \\[0.2cm] 
\mu^c   \propto \overline{f} u + f^+ v , \;& 
\overline{\mu^c} \propto \overline{f}^+ \overline{u} + f \overline{v} . 
\ea
\ee
Here $f$ annihilates a particle and  
$\overline{f}^+$ creates an antiparticle. 
Using all the above equations the matrix element 
describing the $\nnu \! \!$--like process can be written as 
(see Fig.\ \ref{feyndia} for the definition of the kinematics)
\bea \label{matrix1}
\mbox{$\cal{M}$}_1 \propto \lambda_m^{\ast} U_{\mu m}^{\ast \, 2} 
\overline{\mu^c} \gamma_{\rho} \gamma_- \nu_m \overline{\nu_m} 
\gamma_{\pi} \gamma_+ \mu \\[0.2cm] 
= \lambda_m^{\ast} U_{\mu m}^{\ast \, 2} \left[ \overline{u}(k_2) 
\gamma_{\rho} \gamma_- (\Slash {q}_2 + m_m ) 
\gamma_{\pi} \gamma_+ v (k_3) \right] \frac{\D 1}{\D q_2^2 - m_m^2} \\[0.2cm]
= \lambda_m^{\ast} U_{\mu m}^{\ast \, 2} m_m 
 \left[ \overline{u}(k_2)  \gamma_{\rho} \gamma_- \gamma_{\pi} 
 v (k_3) \right]  \frac{\D 1}{\D q_2^2 - m_m^2}.
\eea
From here on we neglect the mass $m_m$ in the denominator. See below for 
the case when this is no longer allowed. 
The above is the matrix element one would have obtained for an intermediate 
Dirac neutrino and applying the usual Feynman rules 
with one outgoing $\mu^+$ written with an $\overline{u}$ instead of a $v$ 
(thus producing a scalar expression) and one 
$\gamma_+$ replaced with a $\gamma_-$. 
Assuming CP invariance, 
the term $\lambda_m^{\ast} U_{\mu m}^{\ast \, 2} m_m$ can be written as 
(see e.\ g.\ \cite{kayser,bilenky})  
\be \label{meff}
\left| \sum_m \lambda_m^{\ast} U_{\mu m}^{\ast \, 2} m_m \right| = 
\left| \sum_m m_m \eta_m^{\rm CP}  U_{\mu m}^2 \right| \equiv 
\langle m_{\mu \mu} \rangle  , 
\ee
thus defining the usual effective mass. 
The matrix element is therefore 
proportional to the effective Majorana mass, just as in 
\nnu and the other mentioned lepton--number violating processes.\\
For the crossed diagram, 
described by $\mbox{$\cal{M}$}_2$,  
$q_2$ is replaced by $\tilde{q}_2$ and 
$k_2$ by $k_3$. Finally, the interference term is given by  
\be
- \mbox{$\cal{M}$}_1^{\ast} \mbox{$\cal{M}$}_2 \propto 
\overline{v}(k_3) \gamma_{\nu} \gamma_+ \gamma_{\mu} u(k_2) 
\overline{u}(k_3) \gamma_{\alpha} \gamma_- \gamma_{\beta} v(k_2)  . 
\ee
which has a negative sign due to the 
interchange of two identical fermion lines. 
Using the identities $\overline{v} = -u^T C^{-1}$ and 
$u = C \overline{v}^T$ this can be written in a form suitable for 
using the completeness relations for the spinors: 
\bea 
\overline{v}(k_3) \gamma_{\nu} \gamma_+ \gamma_{\mu} u(k_2) = 
- u^T(k_3) C^{-1} \gamma_{\nu} \gamma_+ \gamma_{\mu}  C \overline{v}^T (k_2) 
\\[0.2cm] 
= - u^T(k_3) \left( \gamma_{\mu} \gamma_+ \gamma_{\nu} \right)^T 
 \overline{v}^T (k_2) = \overline{v} (k_2) \gamma_{\mu} \gamma_+ 
\gamma_{\nu} u(k_3) . 
\eea
Putting all the couplings and propagators together, the matrix element 
for scattering with a quark
can be written as (using the Feymann--gauge for the $W$ propagator) 
\begin{small}
\begin{center}
\bea \label{matrix} 
|\overline{\mbox{$\cal{M}$}}|^2 = 
|\overline{\mbox{$\cal{M}$}_1}|^2 + 
|\overline{\mbox{$\cal{M}$}_2}|^2 
+ 2 \Re \, 
\left( \overline{\mbox{$\cal{M}$}_1^{\ast} \mbox{$\cal{M}$}_2} \right) 
\\[0.2cm] 
= |\langle m_{\mu \mu} \rangle|^2 64 G_F^4 M_W^8 
\left| \frac{\D 1 }{\D q_1^2 - M_W^2} \frac{\D 1 }{\D q_3^2 - M_W^2}
\right|^2 \\[0.2cm]
\mbox{ Tr} \mbox{\Large\{} \gamma^{\mu} \gamma_- \Slash {p}_1 \gamma^{\beta} 
\gamma_- (\Slash {k}_1 + m_{\mu}) \mbox{\Large\}}
\mbox{ Tr} \mbox{\Large\{} 
\gamma^{\nu} \gamma_- (\Slash {p}_2 + m_q) \gamma^{\alpha} 
\gamma_- (\Slash {k}_4 + m_{q'}) \mbox{\Large\}} \\[0.2cm]
\left[ \left|\frac{\D 1}{\D q_2^2}\right|^2 \mbox{ Tr} 
\mbox{\Large\{} \gamma_{\mu}\gamma_- \gamma_{\nu} (\Slash {k}_3 - m_{\mu}) 
\gamma_{\alpha} \gamma_+ \gamma_{\beta} 
(\Slash {k}_2 + m_{\mu}) \mbox{\Large\}} \right. \\[0.2cm]
\left. + \left|\frac{\D 1}{\D \tilde{q}_2^2}\right|^2 \mbox{ Tr} 
\mbox{\Large\{}\gamma_{\mu}\gamma_- \gamma_{\nu} (\Slash {k}_2 - m_{\mu}) 
\gamma_{\alpha} \gamma_+ \gamma_{\beta} (\Slash {k}_3 + m_{\mu}) 
\mbox{\Large\}} \right. \\[0.2cm] 
\left. - 2 \left| \frac{\D 1}{\D q_2^2} 
\frac{\D 1}{\D \tilde{q}_2^2} \right| \mbox{ Tr}
\mbox{\Large\{} 
\gamma_{\beta}\gamma_+ \gamma_{\alpha} (\Slash {k}_3 + m_{\mu}) 
\gamma_{\mu} \gamma_- \gamma_{\nu} (\Slash {k}_2 - m_{\mu}) 
\mbox{\Large\}} \right]\\[0.2cm]
= |\langle m_{\mu \mu} \rangle|^2 G_F^4 M_W^8 
\left|\frac{\D 1 }{\D q_1^2 - M_W^2}\frac{\D 1 }{\D q_3^2 - M_W^2}\right|^2 
2^{12} (p_1 \cdot p_2) \\[0.3cm] 
\left[
\left|\frac{\D 1}{\D q_2^2}\right|^2 (k_1 \cdot k_2) (k_3 \cdot k_4) + 
\left|\frac{\D 1}{\D \tilde{q}_2^2}\right|^2 
(k_2 \cdot k_4) (k_1 \cdot k_3) \right. \\[0.2cm] 
\left. -  \left|\frac{\D 1}{\D q_2^2} \frac{\D 1}{\D \tilde{q}_2^2}\right| 
\left( (k_2 \cdot k_3) (k_1 \cdot k_4) - 
(k_2 \cdot k_4) (k_1 \cdot k_3) - (k_1 \cdot k_2) (k_3 \cdot k_4) \right) 
\right] . 
\eea 
\end{center}
\end{small}
$m_q$ and $m_{q'}$ are the masses of the scattered initial and 
final state partons, respectively. Coupling to an antiquark 
is identical to replacing $k_4$ with $p_2$. 
The two short traces describe the 
SM V--A vertices, the ones inside the square brackets 
are the $\nnu \! \!$--like process. 
Averaging over the parton spin adds an additional factor 1/2.
The long traces were computed with the MATHEMATICA \cite{Mathematica} 
package TRACER \cite{tracer}. As can be seen, in the low mass regime 
the matrix element is proportional to $\langle m_{\mu \mu} \rangle^2$. 
If we take a heavy Majorana neutrino into account, one has to include the 
mass in the propagator for $q_2$ and $\tilde{q}_2$, which we neglected from  
Eq.\ (\ref{matrix1}) on. 
A statistical factor of 1/2 due to two identical final state muons 
has to be included in order to avoid double counting in the phase 
space integration.\\ 
We also performed the calculation for purely right--handed (RH) 
currents and obtained the exact same result with the exchange $W \to W_R$. 
As known, RH currents must occur --- if they exist --- 
strongly suppressed with respect to the left--handed ones.
Since the $W$ momenta
are relatively small in comparison to $M_W$, the cross section is
proportional to 
\be
\sigma \propto G_F^4 M_W^8 ((q_1^2 - M_W^2) (q_3^2 - M_W^2))^{-2}
\sim G_F^4 \propto \left( \frac{\D g^2}{\D M_W^2} \right)^4 \, ,
\ee
forcing the purely RH case to be some orders of
magnitude under the
purely left--handed case, since $M_{W_R} > 6 M_W$ \cite{PDG}. 
Here we assumed $g_L = g_R = g$.\\ 
One could also consider a heavy right--handed Majorana neutrino  
as suggested by some left--right symmetric theories \cite{LR}, where
leptons are 
arranged symmetrically in left--handed (LH) and RH doublets, i.\ e.\ 
\be \begin{array}{ccc} 
\left( \begin{array}{c} \nu_{\mu} \\ \mu^- \end{array} \right)_{\rm L} & 
\mbox{ and } & 
\left( \begin{array}{c} N_{\mu} \\ \mu^- \end{array} \right)_{\rm R} . 
\end{array}
\ee 
Here $N_{\mu}$ is a heavy Majorana neutrino 
in a weak leptonic current of the form 
\be \label{jright}
j_l^{\alpha} = \overline{\mu} \gamma^{\alpha} \gamma_+ N_{\mu} + 
\overline{\mu} \gamma^{\alpha} \gamma_- \nu_{\mu} + \ldots 
 + {\rm h.\; c.} 
\ee
where the dots denote non muonic contributions. We consider it 
in order to illustrate the general properties
of process (\ref{proces}) in a model independent way and to stress the fact 
that the greatest sensitivity is achieved for a Majorana 
mass of 1 to 10 GeV, independent of the exact form of the 
coupling to the $W$, see below. 
Furthermore it serves as a comparison to the results 
from \cite{halprin,ng}, 
who also considered this possibility. 
In general all possibilities could contribute at the same time.   
Performing 
the same calculation as before we get for the $N_{\mu}$--case 
in Eq.\ (\ref{matrix}) 
a replacement $(\gamma_+ \leftrightarrow \gamma_-)$ for the trace describing 
the $\nnu \! \!$--like process which leads in the end to a replacement 
($k_1 \leftrightarrow p_1 , \; k_4 \leftrightarrow p_2$). 
Again, antiquark scattering is obtained 
by replacing $k_4$ with $p_2$ in the quark amplitude.\\
Also possible is the exchange of other hypothetical particles such as 
those from the plethora of SUSY\@. Anyway, if process (\ref{proces}) 
could be detected, general arguments guarantee a Majorana mass term for the 
muon neutrino, just as the Schechter and Valle argument \cite{scheval} 
does in the case for \onbb for the electron neutrino. 
In \cite{klapsusy} this theorem has been generalized to supersymmetry 
demanding also a non--vanishing Majorana mass for the scalar superpartners 
of the SM neutrinos. 
The smallness of the cross section however makes a more detailled analysis 
in this case not worthwhile: 
One could in principle derive limits on the 
right--handed coupling and/or $W_R$ mass but they would definitely 
not compete with bounds derived by other methods, in contrast to 
the bound we will derive on \mmm{} in the next section.

\section{Results and Discussion}
\begin{figure}[hp]
\setlength{\unitlength}{1cm}
\vspace{-2.2cm}
\begin{center}
\epsfig{file=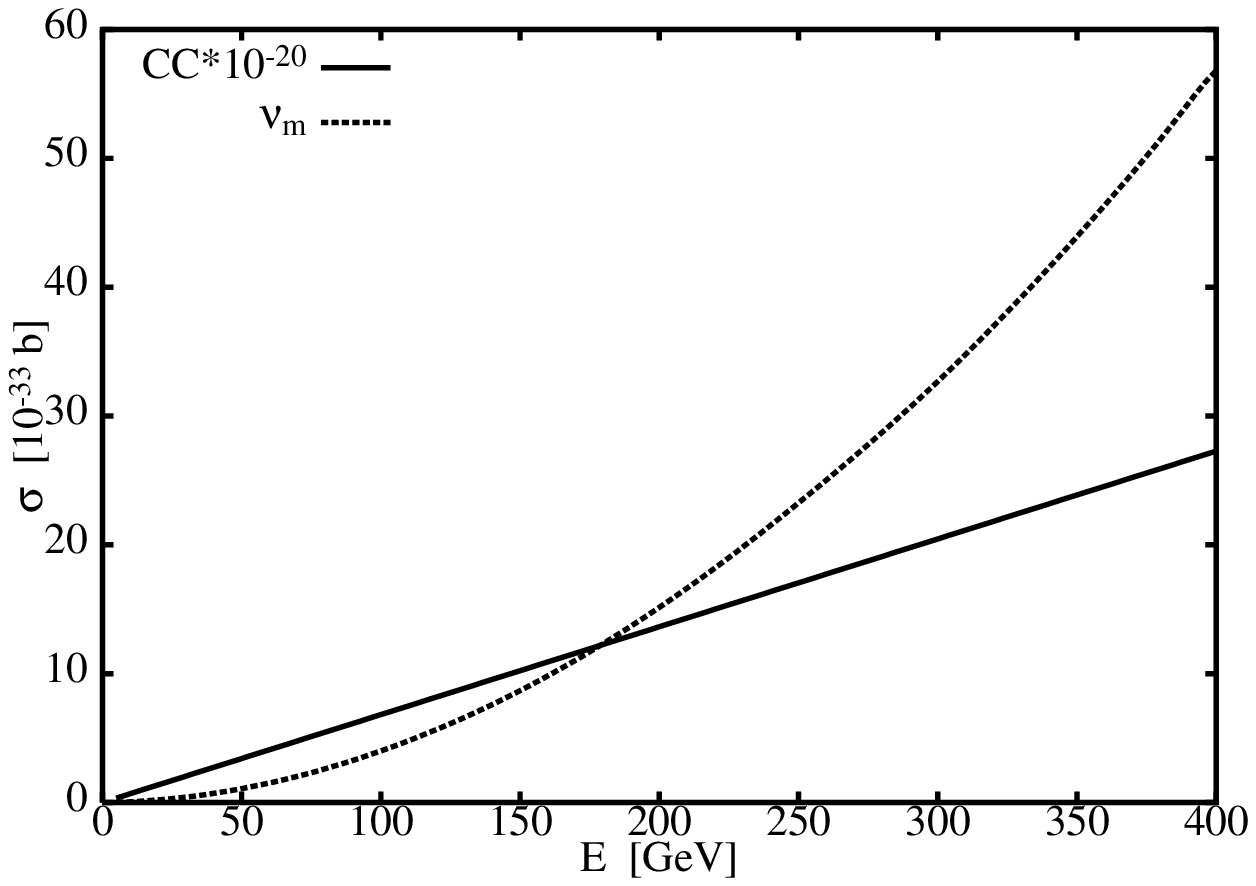,width=13cm,height=8cm}
\end{center}
\vspace{-0.5cm}
\caption{\label{sigvonE}Total cross section for process (\ref{proces}) 
with a left--handed (solid) $\nu_m$ together with the 
total CC $\nu$N cross section 
times $10^{-20}$ (dashed).
The (effective) mass for the neutrino is $\emm \! = 170$ keV\@.}
\vspace{0.5cm}
\setlength{\unitlength}{1cm}
\begin{center}
\epsfig{file=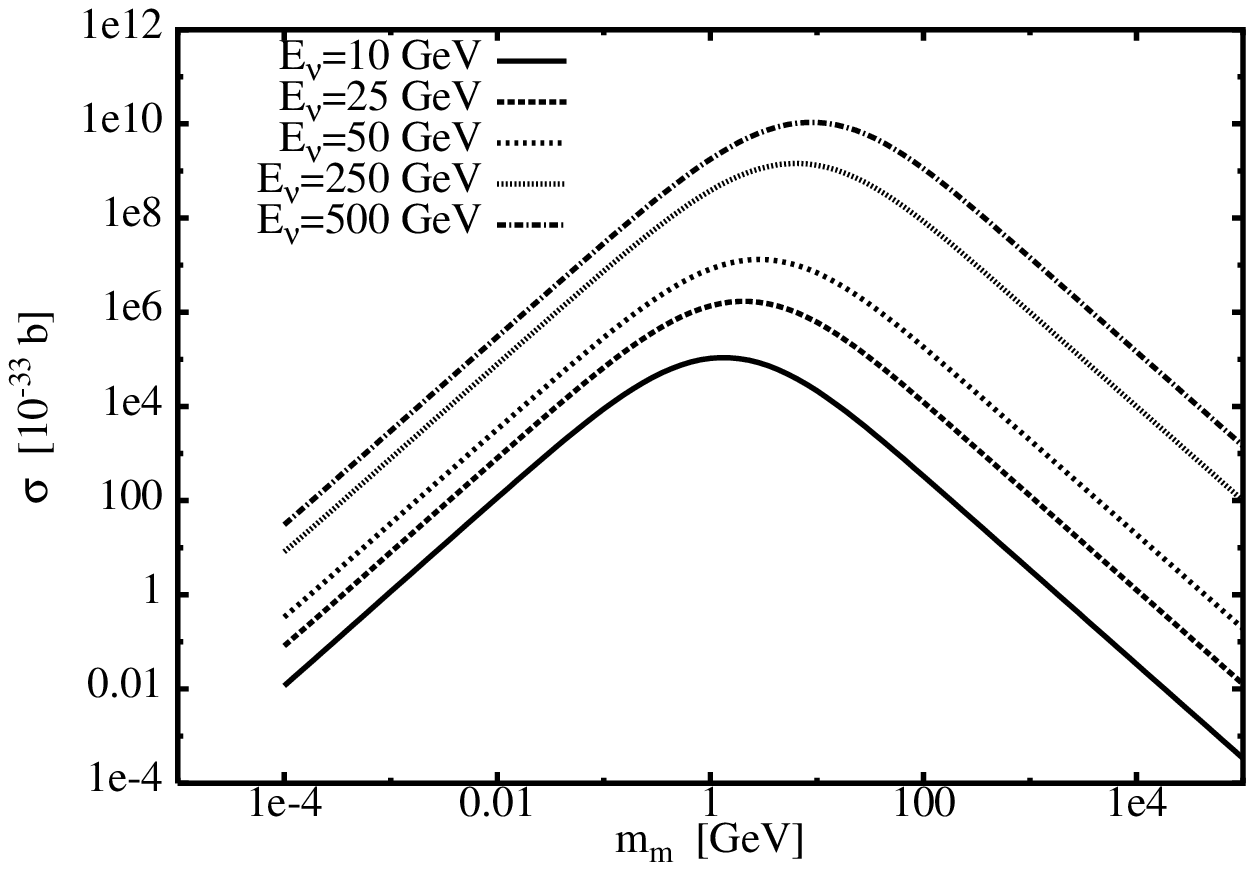,width=13cm,height=8cm}
\end{center}
\vspace{-0.7cm}
\caption{\label{sigvonMl}Total cross section for a left--handed
Majorana neutrino as a function of its mass for different neutrino beam 
energies. No 
limit on $U_{\mu m}^{2}$ was applied. }
\end{figure}

\begin{figure}[hp]
\setlength{\unitlength}{1cm}
\vspace{-2.2cm}
\begin{center}
\epsfig{file=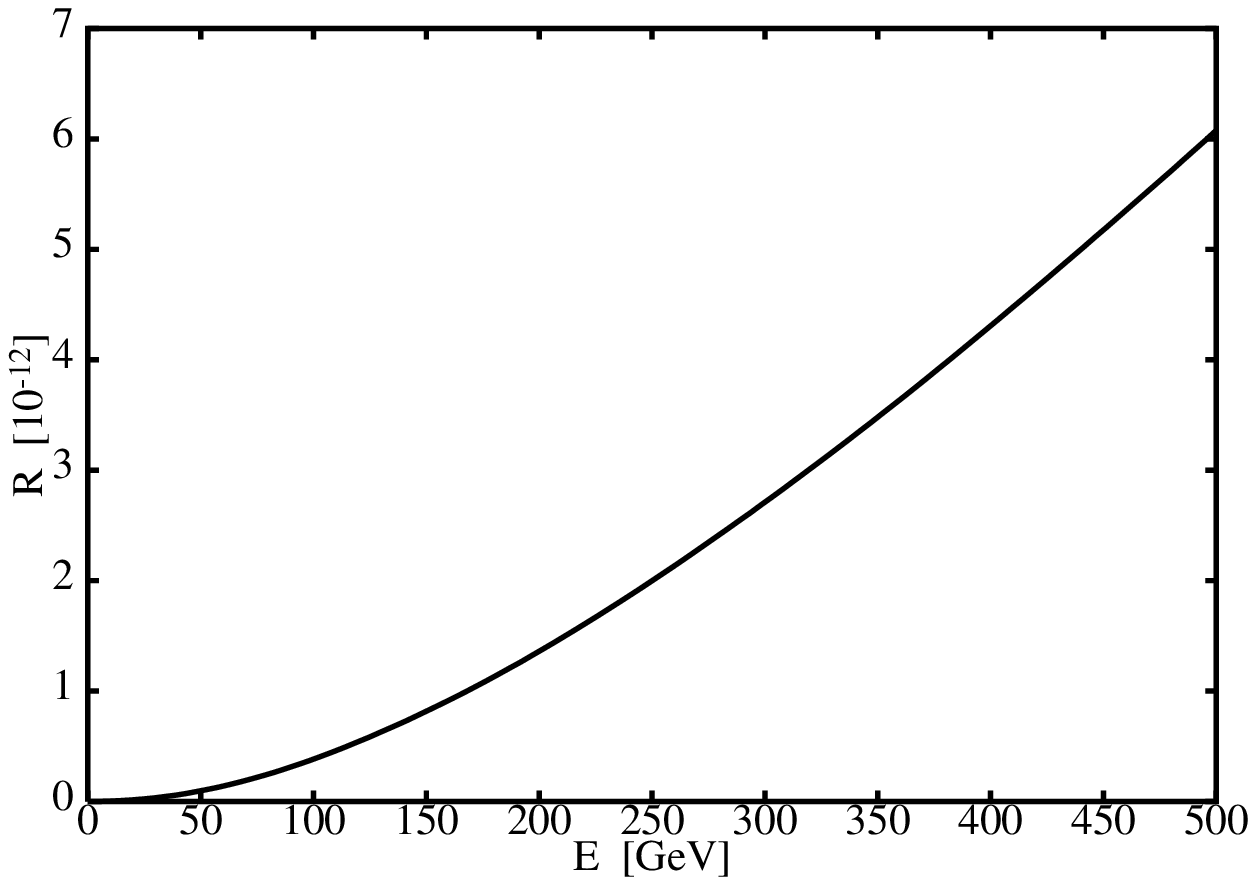,width=13cm,height=8cm}
\end{center}
\vspace{-0.7cm}
\caption{\label{BR}
Ratio for process (\ref{proces}) with respect to 
the total CC $\nu$N cross section for a left--handed Majorana of 
(effective) mass 5 GeV\@. No limit on $U_{\mu m}^{2}$ was applied. }
\vspace{0.1cm}
\setlength{\unitlength}{1cm}
\begin{center}
\epsfig{file=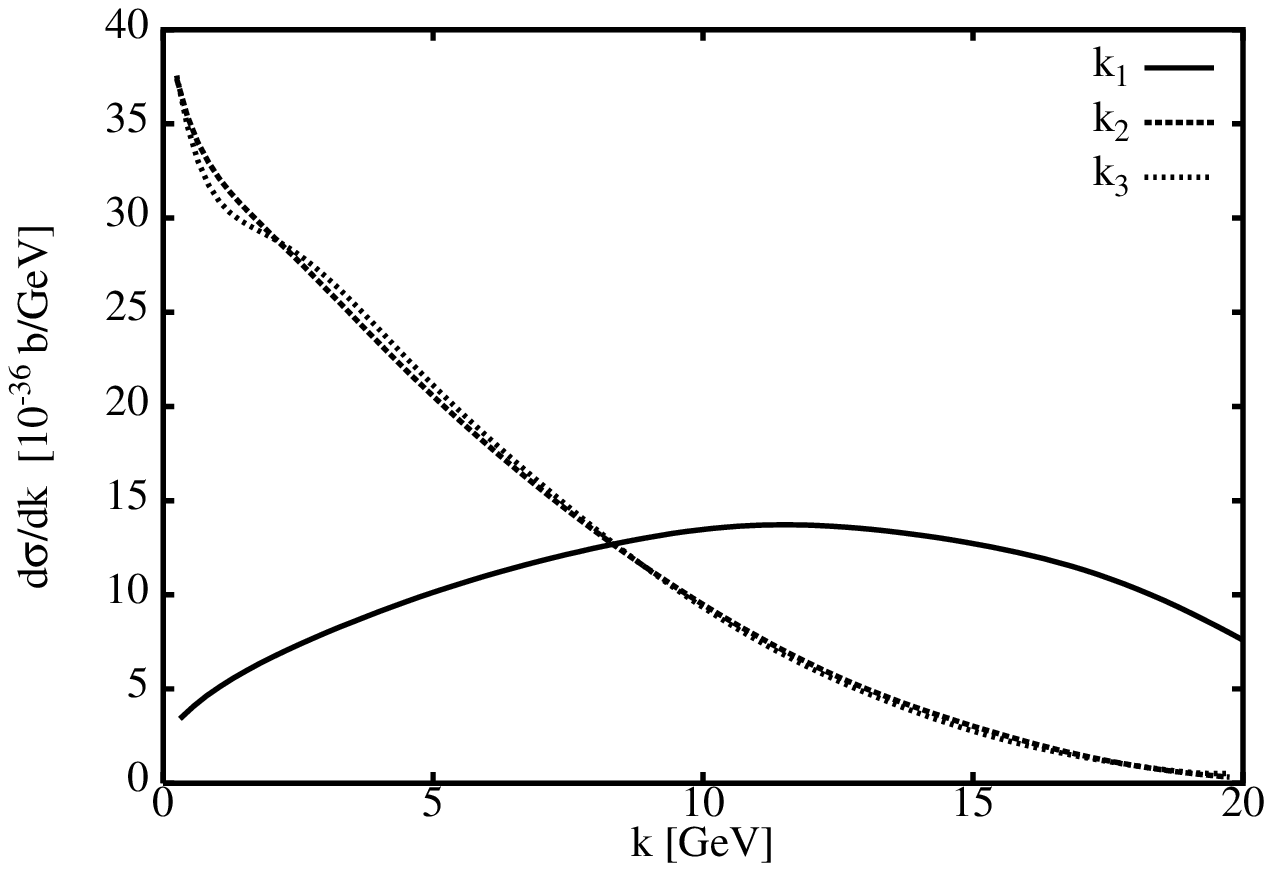,width=13cm,height=8cm}
\end{center}
\vspace{-0.5cm}
\caption{\label{sigp17025l}
Differential cross section at $E_{\nu} = 25$ GeV
for the momenta of the three muons for the case of a 
left--handed Majorana with effective mass of 170 keV\@.
$k_1$ is the muon momentum from
the standard V--A vertex for the incoming neutrino, $k_2$ and $k_3$ 
are the muons from the $\nnu \! \!$--like vertices.}
\end{figure}

For the evaluation of the total and differential cross sections we wrote  
a Monte Carlo program calling the phase space routine GENBOD \cite{genbod}. 
For the parton distributions we chose GRV 98 ($\overline{\rm MS}$) NLO 
\cite{grv98} at $Q^2 = s = (p_1 + p_2)^2 = x^2 M_p^2 + 2 x M_p E_{\nu}$, 
where $M_p$ denotes the proton mass, $E_{\nu}$ the energy of the incoming 
neutrino and $x$ the Bj\o rken variable. 
We set $Q^2 = Q^2_{\rm{min}}$ whenever $Q^2$ went under the 
minimal allowed value of 0.8 GeV$^2$. 
To get the averaged neutrino--nucleon cross section we assumed an isoscalar 
target and replaced up-- and down quarks to get the parton 
distributions for the neutron.\\ 
Before presenting the results we estimate the ratio with 
respect to the total neutrino--nucleon cross section. The typical 
suppression factor one encounters when dealing with 
Majorana instead of Dirac neutrinos is $M/E$ in the matrix elements with 
$M$ being the Majorana's mass and $E$ its energy. 
For the ratio R of the cross sections we have therefore: 
\be
{\rm R} = {\rm \frac{ \sigma (\nmu N \ra \mu^- \mu^+ \mu^+ X)}
{\sigma ( \nmu N \ra \mu^- X)}} \propto 
\left( \frac{M}{E} \right)^2 G_F^2 M_W^4   \sim 
\left\{ 
\begin{array}{cc} 10^{-13}  & 
\mbox{ for } M = 170 \; {\rm keV} \\[0.16cm]
10^{-5}  & 
\mbox{ for } M = 5 \; {\rm GeV} \end{array} \right. \, , 
\ee
where we took as a typical value $E$ = 30 GeV\@. 
For heavy neutrinos the behavior changes significantly: 
Instead of $M/E$ we have now $M^{-2}$ (see below) and the ratio goes as 
\be
{\rm R} \propto \frac{G_F^2 M_W^4 M_p^2}{M^2} \sim 10^{-7} 
\mbox{ for } M = 100 \; {\rm GeV}
\ee
These ratios will of course be further suppressed by a very small 
phase space factor,  
which rises slightly with energy and turns out to be about $10^{-7}$, 
as well as by factors arising from bounds on the mixing with heavy 
neutrinos, see below.\\ 
As expected, the \cs{} is tiny:   
If we use for the mass of the Majorana neutrino the current limit 
from the direct muon neutrino mass search, $m_{\nu_{\mu}} = 170$ 
keV \cite{asa}, 
we find that the cross section for energies in the range $5 \ldots 500$ GeV 
is of the order 
$\sigma (3 \mu) \simeq 1 \ldots 10^2 \cdot 10^{-33} \mbox{ b}$, 
being 20 orders of magnitude lower than the total 
neutrino nucleon CC cross section of 
$ \sigma_{\rm CC} \simeq 1 \ldots 10^2  \cdot 10^{-14} \mbox{ b}  $ 
for the same energy range. 
The $E_{\nu}$ dependence of the cross section can be fitted as 
a quadratically polynom, i.\ e.\ 
$\sigma (3 \mu, E_{\nu})  = a \cdot E_{\nu} + b \cdot E_{\nu}^2 $ 
which has to be compared to the linear dependence of the total CC 
neutrino--nucleon cross section.  
If we assume that this behavior holds up to ultrahigh energies 
(which it does not due to propagator effects) the cross sections 
would be roughly equal for $E_{\nu} \simeq 10^{20}$ GeV, far beyond 
any reasonable scale.\\
The 
scaling with $\langle m_{\mu \mu} \rangle^2$ holds up to masses of 
about 1 to 10 GeV\@. 
The RH $N_{\mu}$ produces a signal in the same order of magnitude. 
We plot the trimuon cross section 
in Fig.\ \ref{sigvonE} 
together with the total CC neutrino nucleon \cs{} of about 
$0.7 \cdot 10^{-14} \, E_{\nu}/\rm GeV$ b, multiplied with $10^{-20}$.\\
Despite the small values, the ratio of 
the trimuon process described here is significantly more 
sensitive on \emm{}than other discussed processes:   
Abad et al.\ \cite{abad} get in a relativistic quark model 
for the decay $K^+ \ra \pi^- \mu^+ \mu^+$ 
a branching ratio of $2 \cdot 10^{-22}$ while Missimer  
et al.\ \cite{moha} estimate the ratio of 
$\mu^-  \mu^+$--conversion via capture in $^{44}$Ti with respect to 
a normal CC reaction to be $4 \cdot 10^{-25}$ for 
a few hundred keV Majorana. Thus the process (\ref{proces}) is 
about two orders of magnitude closer to the relevant standard model process 
than previously discussed Majorana induced muon--number violating processes.\\
Considering now the massive case, i.\ e.\ including the Majorana 
masses in the propagator, the squared amplitude will now be 
proportional to the sum 
\be
\sigma \propto
\left| 
\sum\limits_m \frac{m_m \eta_m^{\rm CP}  U_{\mu m}^2 }
{(q_2^2 - m_m^2)}
\right|^2 .
\ee
For the sake of simplicity we skip for the moment the factors
$\eta_m^{\rm CP}  U_{\mu m}^2$ and consider only one mass eigenvalue, 
which turns out to dominate the cross section when it has an 
appropriate value. 
First of all, the cross section as a function of mass will rise  
quadratically until the propagator takes over and 
forces a (mass)$^{-2}$ behavior.  
This is displayed in Fig.\ \ref{sigvonMl} where we plot the 
total \cs{} for different neutrino energies. As can be seen the 
maximal value of the cross section as a function of mass 
is obtained in the range $1 \ldots 10$ 
GeV, rising slightly with $E_{\nu}$. The reason for that is that the 
integration over the neutrino propagator has its maximum in this range. 
This fact makes the greatest sensitivity 
independent of the coupling of the 
Majoranas to the leptons or $W$'s. One can show  
that the heavy right--handed $N_{\mu}$  
displays the same behavior as the left--handed Majorana case shown in 
Fig.\ \ref{sigvonMl}, which underlines this fact.  
In Fig.\ \ref{BR} we display the ratio with respect to the total CC 
neutrino--nucleon cross section for a mass of 5 GeV 
as a function of the incoming neutrino energy.
Note that this light masses are ruled out \cite{PDG} and that for higher 
masses ($M \gs 100$ GeV) the ratio scales with $M^{-2}$.\\
\begin{table}[t]
\begin{center} 
\begin{tabular}{|c|c|c|c|c|} \hline 
& \multicolumn{2|}{c|}{$m_m$ = 7 GeV}  & 
\multicolumn{2}{c|}  {$m_m$ = 80 GeV} \\ \hline 
 $E_{\nu}$ & R  & $\rm R_{cor}$ & R  & $\rm R_{cor}$ \\ \hline 
25 & $1.1 \cdot 10^{-14}$ & $4.4 \cdot 10^{-24}$ & $1.1 \cdot 10^{-16}$ 
& $4.1 \cdot 10^{-21}$ \\ \hline
50 & $6.2 \cdot 10^{-14}$ & $2.5 \cdot 10^{-23}$ & $8.5 \cdot 10^{-16}$ 
& $3.1 \cdot 10^{-20}$ \\ \hline
100 & $3.1 \cdot 10^{-13}$ & $1.2 \cdot 10^{-22}$ & $6.3 \cdot 10^{-15}$ 
& $2.3 \cdot 10^{-19}$ \\ \hline
250 & $2.1 \cdot 10^{-12}$ & $8.4 \cdot 10^{-22}$ & $8.7 \cdot 10^{-14}$ 
& $3.2 \cdot 10^{-18}$ \\ \hline
500 & $7.4 \cdot 10^{-11}$ & $3.0 \cdot 10^{-20}$ & $6.1 \cdot 10^{-13}$ 
& $2.2 \cdot 10^{-17}$ \\ \hline
\end{tabular} 
\caption{\label{table1}Pure (R) and ``$U_{\mu m}$--corrected'' 
($\rm R_{cor}$) ratios of the process for masses of 7 and 80 GeV 
and different neutrino beam energies in GeV\@.}
\end{center}
\end{table}  
Up to now all the numbers given were for $U_{\mu m}^2 = 1$. 
In this case, a maximum of $7.4 \cdot 10^{-11}$ of the CC cross 
section would be reached for a Majorana with mass of about 7 GeV\@.
A neutrino beam of 500 GeV, coming from a high energy and 
luminous $\mu^+ \mu^-$--collider with $10^{13}$ CC events per year could 
in principle produce a few hundred of such events.\\
However, there exist already strong constraints on the matrix 
elements $U_{\mu m}$ from the data. The DELPHI collaboration \cite{delphi} 
examined the mode $Z \ra \overline{\nu} \nu_m$ and found a limit of 
$|U_{\mu m}|^2 < 2 \cdot 10^{-5}$ for masses up to $m_m \simeq 80$ GeV\@. 
For larger masses analyses of neutrino--quark scattering and other 
processes yield $|U_{\mu m}|^2 < 0.0060$ \cite{Ulimits}. 
This pushes the 
best sensitivity range about a factor of 10 towards higher values.\\
In Table \ref{table1} we show the ratios R with and without 
taking into account the limits given above for different 
energies and for Majorana masses of 7 and 80 GeV\@. 
As can be seen one cannot get closer than at most 
$10^{-17}$, even for a 500 GeV neutrino beam. 
In \cite{cvetic} finite width effects were found to 
increase the cross sections for direct heavy Majorana production  
significantly. However, these effects 
show up for high center--of--mass energies and high masses so 
that in our kinematical and mass sensitivity region these effects should be 
negligible.\\ 
Nevertheless, also in the massive case the improvement compared to 
existing numbers is some orders of magnitude: 
Halprin et al.\ \cite{halprin} 
find a ($U_{\mu m}$--corrected) BR smaller than $3 \cdot 10^{-27}$ for 
$K^+ \ra \pi^- \mu^+ \mu^+$ and $\Sigma^+ \ra \pi^- \mu^+ \mu^+$ 
for a universally coupled 5 GeV heavy neutrino and 
Ng and Kamal \cite{ng} get a ($U_{\mu m}$--corrected) branching  
ratio of $1.3 \cdot 10^{-25}$ for a 2 GeV 
right--handed Majorana coupling to the $W$ as in Eq.\ (\ref{jright}).  
This means, for $E_{\nu} = 100$ (500) GeV and few GeV Majoranas, 
process (\ref{proces}) is up to 5 (7) 
orders of magnitude closer to the standard model CC process than 
previously discussed muon--number violating $\Delta L_{\mu} = 2$ 
processes, which are induced by Majoranas. 
Interestingly the highest BR for the mentioned $K$ 
decay in \cite{halprin} is also in the range of 1 to 10 GeV\@.\\
Though the cross section is probably too small to detect this process in 
the near future, it still allows to set bounds on $\emm \!$. 
Let us assume an upper limit on a process like (\ref{proces}) 
of the order $10^{-5}$ of the standard CC 
process (otherwise it would
have been observed already, see Section \ref{expcon})  
and take an energy of $E_{\nu}$ = 100 GeV\@. 
Starting at small masses, i.\ e.\ $\sigma \propto \emm \!\!^2$,  
we find $\emm \! \ls 10^4$ GeV\@. 
This has to be compared to \emm $\! < 1.3 \cdot 10^5$ GeV
as obtained from K--decays \cite{japaner}.\\  
What is now the significance of this \mmm{} bound? 
Obviously, in a three--neutrino framework all elements of \mab{}
should be roughly in the same order of magnitude, therefore at most 
a few eV, not much higher than the limit for \mee{}. 
In scenarios with additional massive neutrinos one has to include 
other processes as restrictions for possible mass matrix models. 
The bounds for 
mixing matrix elements with heavy neutrinos are typically in the order of 
a few $10^{-3}$ to $10^{-2}$ for $\nu_e , \nu_{\mu} \mbox{ and } 
\nu_{\tau}$, thus roughly 
the same for all three families. It is impossible to reconcile the 
bounds for \mee{}, \meu{} and \mmm{} with these conditions. 
If one considers heavy neutrinos and allows \mee{} to be as high as the other 
two one can in principle fulfill the other conditions, but one is lead 
to contradictions when taking flavor changing neutral current processes 
into account, see the Appendix. 
It is thus not possible to construct a mass 
matrix \mab{} spanning 14 orders of magnitude from \mee{} to \mmm{}. 
Therefore, indirect bounds will be far more effective than direct ones. 

\section{\label{expcon}Experimental considerations}
\begin{figure}[hp]
\setlength{\unitlength}{1cm}
\vspace{-2.2cm}
\begin{center}
\epsfig{file=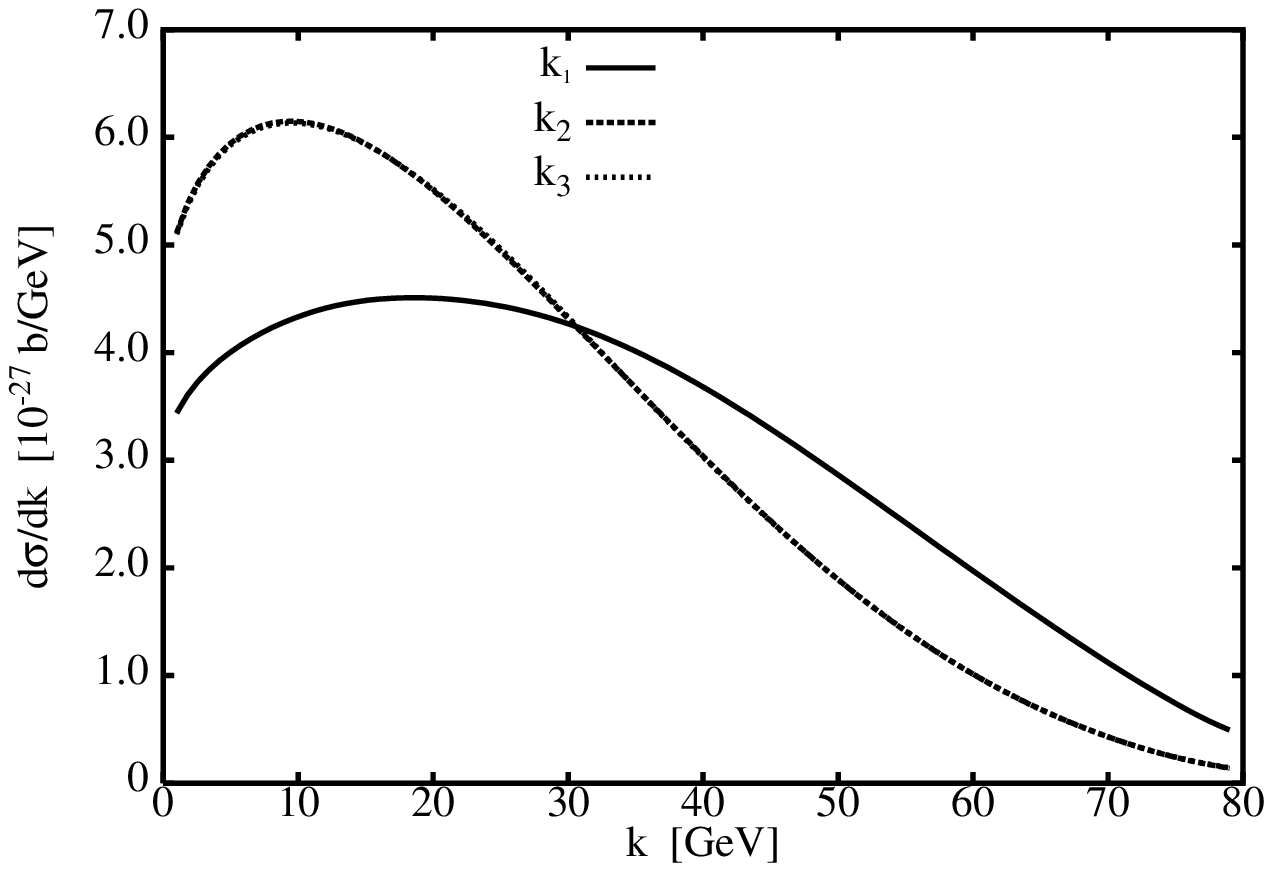,width=13cm,height=8cm}
\end{center}  
\vspace{-0.7cm}
\caption{\label{sigp5100l}Same as previous figure for a mass of 
5 GeV and incoming neutrino energy of 100 GeV\@. 
No limit on $U_{\mu m}^{2}$ was applied.}
\vspace{0.5cm}
\setlength{\unitlength}{1cm}
\begin{center}
\epsfig{file=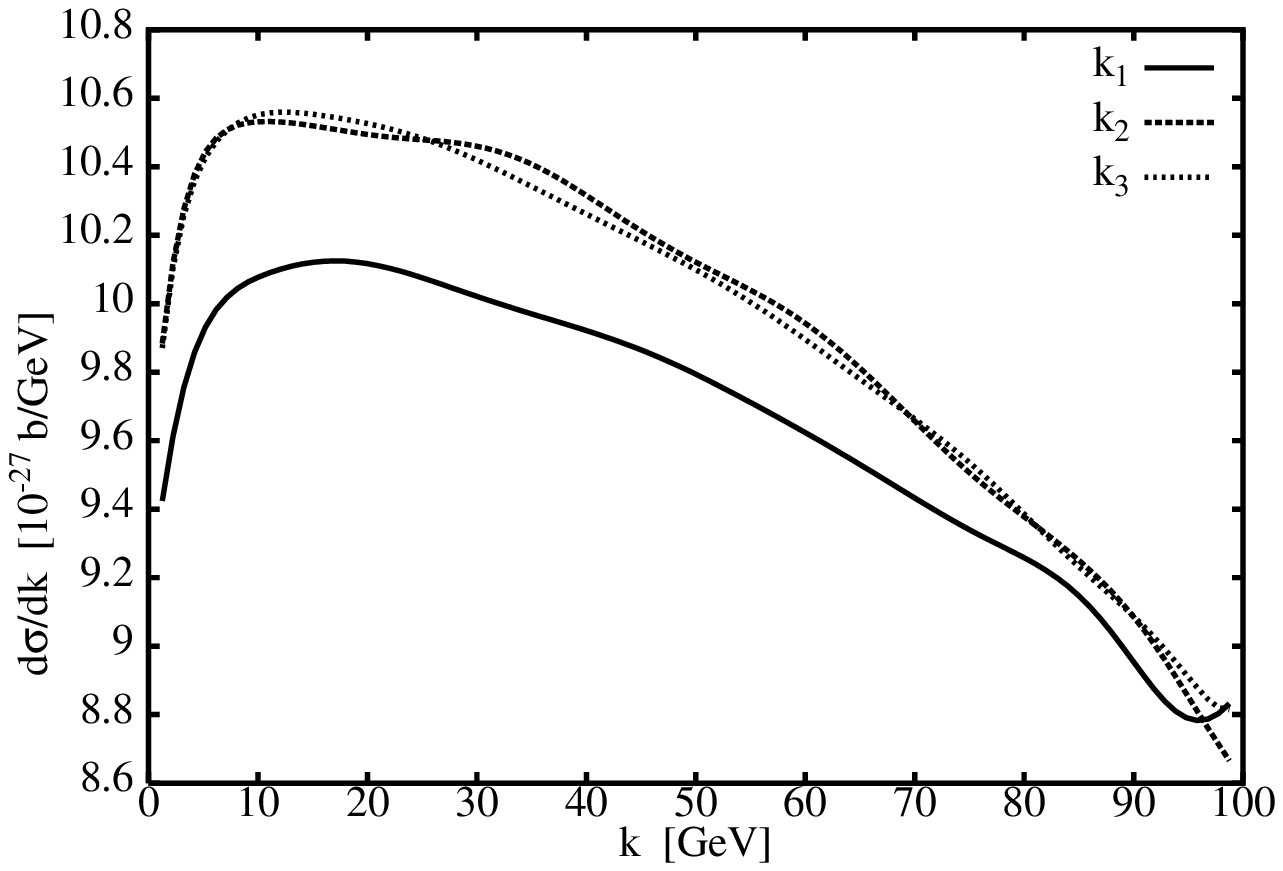,width=13cm,height=8cm}
\end{center}  
\vspace{-0.5cm}
\caption{\label{sigp80500l}Same as above for a mass of 80 
GeV and incoming neutrino energy of 500 GeV\@. Note that in this case the 
momenta of the two $\mu^+$ are always larger than the momentum of 
the $\mu^-$. No limit on $U_{\mu m}^{2}$ was applied.}   
\end{figure}

Several experiments report the observation of trimuon 
events \cite{citf,fhp,cdhs}. 
The observed ratio of trimuon
events (having a lepton number conserving (-- -- +) signature) 
with respect to single
charged current events is of the order $10^{-5}$.
First thought to provide evidence for physics beyond the SM 
the explanation was soon given in terms of CC reactions 
with dimuon production via meson decay, radiative processes 
or direct muon pair production from subsequent hadronic interactions 
\cite{bar78,smi78,barn78}. 
An acceptance cut on muon momenta to be larger than about 
5 GeV was applied by all experiments.\\ 
To extract a (-- + +) signature several background processes in 
typical wide band neutrino beams
have to be considered. Among them the most severe are lepton pair
creation due to antineutrino contaminations of the beam 
(also having a (-- + +) signature) and 
charm production with an associated pion or kaon decay as well as 
overlaying events with beam muons.
Furthermore, going to high momenta some misidentification 
in the charge might lead to additional background.\\
The observables found in the past to be 
most suitable for distinguishing the mentioned standard processes from 
new physics were the momenta of the muons, their two-- and 
three--body invariant masses and the azimuthal angle distribution
between the leading muon and the other two. 
The leading muon ($\mu_1$) was defined as the one which 
minimizes the sum of the transverse momenta of the remaining two 
with respect to the direction of the 
$W, \; ( \vec{W} = \vec{\nu} - \vec{\mu}_1)$. 
A complete listing of all relevant distributions is not our aim, 
however, for the sake of completeness, we plot the distributions of the
muon momenta, which might be used to identify the process.
From Figs.\ \ref{sigp17025l} to \ref{sigp80500l} it can be seen, 
that for the light $\emm \! \!$--case the two $\mu^+$ have 
relatively low energy, while the
$\mu^-$ from the V--A vertex has a broad spectrum with significantly higher
energy. This is no longer valid for a heavy Majorana where the
difference of the muon momenta is less clear,
but is becoming larger with increasing neutrino energy. However, the 
like--sign muons have typically the same momentum distributions, 
which is an important experimental signature.  
It is a general feature that the momentum difference gets 
bigger when the energy $E_{\nu}$ is significantly
higher than the mass of the intermediate 
Majorana. For mass and energy being equal the distributions are more or less 
identical.\\
A similar search to the one described here could also be done 
with $\bnmu$ beams looking for
the corresponding 
process $\bnmu {\rm N} \ra \mu^+ \mu^- \mu^- X$, which though 
turn out to have a cross section lower by a factor of 9 to 10.

\section{Summary and conclusion} 
We investigated the reaction 
$\nu_{\mu} N \ra \mu^- \mu^+ \mu^+ X$ at fixed
target experiments mediated by light and heavy Majorana neutrinos. 
Using the fact, that no excess events were observed in past
experiments at the level of $10^{-5}$ with respect to charged current events, 
we could deduce a limit of $\emm \! \! \ls 10^{4}$ GeV\@. This is 
more stringent than other {\it direct} results discussed on this quantity, 
but obviously not reconcilable with other laboratory experiments.\\
Some general properties of process (\ref{proces}) were discussed:   
The largest sensitivity was found for heavy Majorana neutrinos in the 
region between 1 and 10 GeV because of the fixed target kinematics. 
This was pushed towards approximately 100 GeV due to existing 
limits on $U_{\mu m}^{2}$.   
This is relatively independent of incoming neutrino energy and independent
on the precise form of the couplings, as can be shown with a right--handed 
Majorana. 
In general, process (\ref{proces}) is closer to the standard model CC 
process by 2 (up to a few 100 keV mass) 
up to 7 ($>$ 1 GeV mass) orders of magnitude 
than previously discussed Majorana induced 
$\Delta L_{\mu} = 2$ processes. 
We state again that our $10^4$ GeV 
is the best {\it direct} limit, not the best achievable limit.\\ 
One could consider various modifications of process (\ref{proces}) 
in order to constrain non--standard model 
parameters connected with the muonic sector.  
For the case of $\nnu \!$  
limits on some Yukawa couplings $\lambda_{1jk}^{(')}$, describing 
R--Parity violating SUSY effects were deduced \cite{lam1jk}, the bounds being 
up to four orders of magnitude more stringent than the ones obtained 
from other processes.  
In addition, for muon capture in $^{44}$Ti, extensions of the 
standard model were found \cite{moha} to have branching ratios 
some orders of magnitude higher than the Majorana case, so that it 
seems worthwhile to apply them to process (\ref{proces}) as well.\\
The smallness of the cross section however makes such a detailled 
analysis not very worthwhile. This might change for the 
case of a neutrino factory from a muon storage ring with a large number 
of interactions. Work about this topic is in progress and 
will be presented in the near future.\\
We concentrated our 
analysis on neutrino beams, especially $\nu_{\mu}$. 
Since the beam energies are much higher than the lepton masses, 
the same arguments as described here would hold for other 
fixed--target experiments using charged lepton beams.
However, new background processes have to be considered here.\\
Furthermore, also a lepton--hadron collider 
such as HERA, which also has the 
advantage of higher $\sqrt{s}$ can be used. 
The same strategy that lead to the bound on 
$\langle m_{\mu \mu} \rangle$ can of course be applied to infer 
quantities as $\langle m_{\mu \tau} \rangle$ or 
$\langle m_{\tau \tau} \rangle$, for which no direct 
limits whatsoever exist. Taking the appropriate channels, no SM 
processes faking the signal exist. 
This has in the meantime been discussed in detail in \cite{frz}. 

\hspace{3cm}
\begin{center}
{\bf \large Acknowledgments}
\end{center}
This work has been supported in part (M.\ F.\ and W.\ R.) by the 
``Bundesministerium f\"ur Bildung, Wissenschaft, Forschung und Technologie, 
Bonn under contract number 05HT9PEA5.  
A scholarship (W.\ R.) of the Graduate College 
``Erzeugung und Zerf$\ddot{\rm a}$lle von
Elementarteilchen'' at Dortmund university is gratefully acknowledged.

\setcounter{section}{0}
\renewcommand{\thesection}{\Alph{section}}
\def\chaptername{Appendix}

\section{Appendix}
The task is to construct a mass matrix with $\mee{} \simeq 0$, 
$\meu{} \simeq 2$ GeV and $\mmm{} \simeq 10^{4}$ GeV\@. 
For the matrix elements 
for mixing with heavy neutrinos holds \cite{Ulimits} 
\be \label{ulim}
\sum |U_{e H}|^2 < 6.6 \cdot 10^{-3} \mbox{ and }
\sum|U_{\mu H}|^2 < 6.0 \cdot 10^{-3} . 
\ee
First, let us assume that $U_{eH}^2 \simeq 0$ and 
$U_{ \mu H}^2 = 6 \cdot 10^{-3}$. 
Then we find that $m_H = 1.7 \cdot 10^{6}$ GeV, leading to 
$U_{eH} \simeq 1.6 \cdot 10^{-5}$ which in turn leads to a contribution 
to \mee{} of $4.1 \cdot 10^{5}$ eV, in contradiction to our assumption. 
In general we found no solution for the allowed parameters.\\
Conversely, we might want to use the fact that there is a bound 
on heavy ($m_H > 1$ GeV) contributions from the Heidelberg--Moscow 
experiment of \cite{heavy}
\be \label{uheavy}
\sum U_{eH_i}^2 \frac{1}{m_{H_i}} < 5 \cdot 10^{-5} \mbox{ TeV}^{-1}. 
\ee
Ignoring the condition $\mee{} \simeq 0$ allows to find parameters capable 
of obeying the \meu{} and \mmm{} limits as well as 
Eqs.\ (\ref{ulim}) and (\ref{uheavy}). For example, 
$U_{eH_1}^2 = U_{eH_2}^2 = 10^{-10} , \,  U_{eH_3}^2 = 4 \cdot 10^{-13}$ and 
$U_{\mu H_1}^2 = 5 \cdot 10^{-3} , \, U_{\mu H_2}^2 = -5 \cdot 10^{-4}  
, \,  U_{eH_3}^2 = 10^{-5}$ with $m_{H_1} = 100 \, {\rm GeV} , \, 
 m_{H_2} = 1 \, {\rm TeV} \mbox{ and } m_{H_3} = 10^6 \, {\rm TeV}$.\\ 
Then again we have FCNC processes like $\mu \ra e \gamma$, which are 
sensitive on $\omem = \sqrt{ \sum U_{\mu i} U_{e i} m_i^2}$. The experimental 
value of the branching ratio, ${\rm BR} < 1.2 \cdot 10^{-11}$ \cite{muegamma} 
and the theoretical value from \cite{mohapal} gives  
$\omem < 1.23$ GeV\@, which is not fulfilled by the choice given.

\newpage

\newpage

\newpage

\newpage

\end{document}